\documentclass[a4paper,11pt,fleqn]{article}
\usepackage{authblk}
\usepackage{datetime}

\usepackage{pstricks}
\usepackage{colortbl}
\usepackage{enumitem}

\usepackage{pst-node}
\usepackage{pst-grad}
\usepackage{pst-coil}
\usepackage{pst-text}
\usepackage{pst-rel-points}
\usepackage{amssymb}
\usepackage{amsmath}
\usepackage{mathptm}
\usepackage{fancyvrb}
\usepackage{calc}
\usepackage{etex}
\usepackage{ifthen}
\usepackage{multirow}
\usepackage{rotating}
\usepackage{longtable}
\usepackage[normalem]{ulem}
\usepackage{stmaryrd}
\usepackage{hyperref}
\usepackage{xspace}
\usepackage{mathtools}

\newcommand{\Start}[1]{\text{{\sf\em StartNode}$_{#1}$}\xspace}
\newdateformat{titledate}{\THEDAY\ \monthname[\THEMONTH], \THEYEAR}
\newdateformat{mydate}{\monthname[\THEMONTH] \THEYEAR}

\psset{unit=1mm}

\protect{\newlength{\BRACEL}}
\protect{\newlength{\TAL}} 
\newlength{\codeLineLength}

\newcommand{\buf}{\text{\sf\em buf\/}\xspace}

\newcommand{\bpt}{\text{\sf\em bpt\/}\xspace}
\newcommand{\adv}{\text{\sf\em shift\_offset\/}\xspace}
\newcommand{\str}{\text{\sf\em string\/}\xspace}
\newcommand{\malloc}{\text{\sf\em malloc\/}\xspace}
\newcommand{\calloc}{\text{\sf\em calloc\/}\xspace}
\newcommand{\memcpy}{\text{\sf\em memcpy\/}\xspace}
\newcommand{\strcat}{\text{\sf\em strcat\/}\xspace}
\newcommand{\strncat}{\text{\sf\em strncat\/}\xspace}
\newcommand{\strcpy}{\text{\sf\em strcpy\/}\xspace}
\newcommand{\strncpy}{\text{\sf\em strncpy\/}\xspace}
\newcommand{\strlen}{\text{\sf\em strlen\/}\xspace}

\newcommand{\realloc}{\text{\sf\em realloc\/}\xspace}
\newcommand{\free}{\text{\sf\em free\/}\xspace}
\newcommand{\size}[1]{\text{\sf\em size\text{$#1$}\/}\xspace}
\newcommand{\nps}{\text{\sf\em nps\/}\xspace}
\newcommand{\rnps}{\text{\sf\em rnps\/}\xspace}
\newcommand{\saturate}{\text{\sf\em sat\/}\xspace}
\newcommand{\relop}{\text{\pscirclebox[framesep=.1,linewidth=.15]{\scriptsize\bf?}}}
\newcommand{\pivot}{\text{\sf\em pivot\/}\xspace}
\newcommand{\start}{\text{\sf\em start\/}\xspace}
\newcommand{\strend}{\text{\sf\em end\/}\xspace}
\newcommand{\pointee}{\text{\sf\em pt\/}\xspace}
\newcommand{\shift}{\text{\sf\em shift\/}\xspace}
\newcommand{\glb}{\text{\sf\em glb\_le\/}\xspace}

\newcommand{\bm}{\text{$\alpha$}\xspace}
\newcommand{\bp}{\text{$\beta$}\xspace}
\newcommand{\updatemaps}{\text{\sf\em update\_maps\/}\xspace}
\newcommand{\updatebuf}{\text{\sf\em update\_buf\/}\xspace}
\newcommand{\updatebpt}{\text{\sf\em update\_bpt\/}\xspace}
\newcommand{\npfromy}{\text{\sf\em np\_src\/}\xspace}
\newcommand{\npsfromy}{\text{\sf\em nps\_src\/}\xspace}
\newcommand{\npsbefore}{\text{\sf\em nps\_before\/}\xspace}
\newcommand{\npsafter}{\text{\sf\em nps\_after\/}\xspace}
\newcommand{\npat}{\text{\sf\em np\_at\/}\xspace}
\newcommand{\sdist}{\text{\sf\em shift\_dist\/}\xspace}
\newcommand{\nullpos}{\text{\sf\em length\/}\xspace}
\newcommand{\copypos}{\text{\sf\em copy\_pos\/}\xspace}
\newcommand{\oflow}{\text{\sf\em o\_flow\/}\xspace}
\renewcommand{\inf}{\text{\sf\em inf\/}\xspace}
\newcommand{\fin}{\text{\sf\em fin\/}\xspace}

\newcommand{\aM}{\text{$\sqcap_A\,$}\xspace}
\newcommand{\aO}{\text{$\sqsubseteq_A\,$}\xspace}
\newcommand{\abM}{\text{$\sqcap_{AB}\,$}\xspace}
\newcommand{\abO}{\text{$\sqsubseteq_{AB}\,$}\xspace}
\newcommand{\bM}{\text{$\sqcap_B\,$}\xspace}
\newcommand{\bO}{\text{$\sqsubseteq_B\,$}\xspace}
\newcommand{\bidM}{\text{$\sqcap_N\,$}\xspace}

\newcommand{\nullposM}{\text{$\sqcap_{Z}\,$}\xspace}
\newcommand{\nullposO}{\text{$\sqsubseteq_{Z}\,$}\xspace}
\newcommand{\sizeM}{\text{$\sqcap_S\,$}\xspace}
\newcommand{\sizeO}{\text{$\sqsubseteq_S\,$}\xspace}

\renewcommand{\int}{\text{\sffamily I$^+$}\xspace}
\newcommand{\nds}{\text{\sffamily N}\xspace}
\newcommand{\pts}{\text{\sffamily P}\xspace}
\newcommand{\biinfL}{\text{{\sffamily N}$_b^\infty$}\xspace}
\newcommand{\biL}{\text{{\sffamily N}$_b$}\xspace}
\newcommand{\nullposL}{\text{\sffamily Z$_b$}\xspace}
\newcommand{\sizeL}{\text{\sffamily S$_b$}\xspace}

\newcommand{\bin}{\text{\sf\em bIn\/}\xspace}
\newcommand{\bout}{\text{\sf\em bOut\/}\xspace}

\begin{document}

\date{}

\title{\textbf{Buffer Overflow Analysis for C}}
\author{{\Large Uday P. Khedker}}
\affil{Department of Computer Science and Engineering \\ Indian Institute of Technology Bombay \\ 
Email: \tt {uday@cse.iitb.ac.in}}
\titledate
\maketitle
\thispagestyle{empty}

\begin{abstract}
Buffer overflow detection and mitigation for C programs has been an important concern for a long time. 
This paper defines a string buffer overflow analysis for C. The
key ideas of our formulation are $(a)$~separating buffers from the pointers that point to them, 
$(b)$~modelling buffers in terms of sizes
and sets of positions of null characters, and $(c)$~defining stateless functions to
compute the sets of null positions and mappings between buffers and pointers.

This exercise has been carried out to test the feasibility of describing
such an analysis in terms of lattice valued functions and relations to
facilitate automatic construction of an analyser without the user having
to write C/C++/Java code. This is facilitated by avoiding stateful formulations
because they combine effects through side effects in states
raising a natural requirement of C/C++/Java code to be written to describe them.
Given the above motivation, the focus of this
paper is not to build good static approximations for buffer overflow
analysis but to show how given static approximations could be
formalized in terms of stateless formulations so that they become 
amenable to automatic construction of analysers.

\end{abstract}

\section{Introduction}

Low level modelling of strings in C and associated unchecked operations
potentially lead to the possibility of buffer overflows. Given the
possibility of a potentially fraudulent use of these loop holes in C
programs, detection and mitigation of buffer overflows is 
critical~\cite{%
citeulike:6345464,%
buf.study,%
Lhee:2003:BOF:781669.781672,%
Shahriar:2010:MBO:2441114.2441116,%
Zitser:2004:TSA:1029894.1029911%
}.

This paper proposes an analysis to discover buffer overflows. The 
key ideas of our formulation are $(a)$~separating buffers from the pointers that point to them, 
$(b)$~modelling buffers in terms of sizes
and sets of positions of null characters, and $(c)$~defining stateless functions to
compute the sets of null positions and mappings between buffers and pointer.
The first idea is not new; the novelty of our work lies in modelling the 
computations of null position sets and an insistence on stateless formulations.
As is customary, we present our formulation in an intraprocedural setting. 
It should be easy to lift it to an interprocedural setting using standard techniques of
interprocedural data flow analysis such as the method of value contexts~\cite{%
Khedker.UP.Karkare.B:2008:Efficiency-Precision-Simplicity,%
Khedker.UP.Sanyal.A.Karkare.B:2009:Data-Flow-Analysis,%
Padhye:2013:IDF:2487568.2487569}.

Our goal is not to device the best possible static approximations for buffer
overflow analysis but to show how given static approximations
could be formalized to devise a mathematical formulation which can be
transcribed into a declarative specification of the analysis so that it
becomes amenable to automatic construction of analysers.


The rest of the paper is organized as follows:
Section~\ref{sec:requirements} describes the requirements
of buffer overflow analysis by defining the core statements for analysis, 
the soundness criterion, and our assumptions. 
Section~\ref{sec:formulation} describes our modelling and defines the
analysis in terms of lattices and lattice valued functions and relations. For 
simplicity of exposition, it assumes that a pointer points to
a single buffer at a program point.
Section~\ref{sec:exmp} shows the analysis of our running
example.  
Section~\ref{sec:extensions} shows how the model can be extended to allow
a pointer to point to multiple buffers at a program point.
Section~\ref{sec:related} briefly describes the related work to highlight the trends while
Section~\ref{sec:conclusions} concludes the paper.
Appendix~\ref{app:stmt.modelling} describes how other statements can be modelled
in terms of core statements.

\section{Requirements of Buffer Overflow Analysis}
\label{sec:requirements} 

Our formulation is guided by the following requirements and assumptions.

\subsection{Program Model}

We assume the following model of programs.
\begin{itemize}
\item A buffer is an array of \verb+char+ or pointers to \verb+char+
      storing C-style strings. It could contain multiple strings and
      hence multiple null characters (\verb+'\0'+). A pointer may point
      to any location within a buffer.
\item The list of core buffer manipulation statements to be considered for this analysis is as follows. 

\begin{itemize}
\item Buffer creation: \text{$x = \malloc(i)$} and \text{$x =
      \calloc(i)$} where $i$ is a compile time constant, and
      \text{$\free(x)$}.

\item Buffer assignment: 
	\begin{itemize}
	\item \text{$x = y$} and \text{$x = y+i$} where $i$ is a compile time constant.
	\item \text{$\memcpy(x,y,m)$} which copies a $m$ character long block of memory
               pointed to by $y$ to the memory pointed to by $x$.
	\end{itemize}
\item Buffer modification:
	\begin{itemize}[label=o]
        \item Direct modification. \text{$x[i] =\, '\!\backslash{}0'$}
              and \text{$x[i] =\, '\!c'$} where $i$ is a compile
              time constant and c is a non-null character.

	\item Modification through string functions.
	\text{$\strcat(x,y)$}
	\text{$\strcpy(x,y)$} and their length limited versions 
	(\text{$\strncat(x,y,m)$} and \text{$\strncpy(x,y,m)$}).
	\end{itemize}
\item Buffer reading: Any statement using $x$,
	\text{$x[i]$}, or
	\text{$*(x+i)$} as an rvalue or calls to 
	\text{$\strlen(x)$}. 
\end{itemize}
\item A program is viewed as a convention control flow graph with each node
      representing a single statement.
\end{itemize}

Appendix~\ref{app:stmt.modelling} explains how
other statements are modelled in terms of these statements.

\subsection{Soundness Criterion,  Required Approximations, and Assumptions}
\label{sec:assumptions}
We assume the following soundness criterion: no buffer overflow should
go undetected; false positives about buffer overflow can be tolerated.

In order the ensure soundness, we introduce the following approximation:
      A buffer has a single set of null positions associated with it. If the
      sets of a buffer differ along different execution paths reaching a
      program point, we create an approximate buffer such that regardless of
      the execution path, every string contained in the original buffer is a
      substring of some string present in the approximate buffer.
This approximation may cause some imprecision in that our analysis
may consider longer strings than are actually present in the
buffer leading to false positives.

This approximation is implemented in the following manner:
     \begin{itemize}
     \item At a given program point, a buffer could have different null position
           sets along different control flow paths reaching the program point. 
           Hence, at the join points in the program, the null position sets of
           a buffer reaching along different control flow paths are intersected.

	\item We assume that the memory allocated using \malloc does not contain a null character.

	\item  We assume that the values of integer variables appearing in 
               \text{$\malloc(i)$}, \text{$x[i]$}, or \text{$x = y + i$} are
               known at compile time or a range analysis has been performed.  
               If range information is available, we choose the low limit of the range
               of $i$ for \text{$\malloc(i)$} and the 
               the high limit of the range of $i$
              for \text{$x[i]$} and \text{$x = y + i$}.
		If $i$ is not a compile time constant and its range is not known, 
               we assume its value to be $\infty$.
              
        \end{itemize}

\section{Formulating Buffer Overflow Analysis}
\label{sec:formulation}

In this section, we model buffers and the relations of pointers holding the 
addresses of buffers. This is then followed by  formulating data flow equations that compute them.
As is customary, we present our formulation in an intraprocedural setting. 
It should be easy to lift it to an interprocedural setting using standard techniques of
interprocedural data flow analysis such as the method of value contexts~\cite{%
Khedker.UP.Karkare.B:2008:Efficiency-Precision-Simplicity,%
Khedker.UP.Sanyal.A.Karkare.B:2009:Data-Flow-Analysis,%
Padhye:2013:IDF:2487568.2487569}.

\subsection{Assumptions for Simplifying the Formulation}

For simplicity of exposition, we make the following assumptions for the purpose of presenting the
formulation.  Section~\ref{sec:extensions} extends the formulation
by relaxing these assumptions.
     \begin{itemize}
     \item At a given program point, in general, a pointer could point to 
           different buffers along different control flow paths reaching the program point.
           For simplicity of modelling, we assume that a pointer points to a single buffer.
  
        \item 
             Under the assumption that range information is available, we ignore all back edges in the program and
              view it as a directed acyclic graph. This allows us to handle the situation where pointers 
              to a buffer are advanced in a loop and hence at the loop entry, such a pointer points to two different
              positions in a buffer. If we do not ignore back edges, such common case usage of
              pointers will lead to over-approximation of null position sets leading to proliferation of false negatives.

              Loops basically contribute to advancement of pointers or increments of indices. Range information captures
              these effect. Hence we can reduce false negatives by restricting the merge points to those
              resulting from forward edges in the control flow graph of the program. 
        \end{itemize}

Note that a single pointee assumption does not preclude the possibility that a buffer may be pointed to by multiple pointers.
Such a situation is easily handled by our formulation. Indeed, 
our running example of Figure~\ref{fig:exmp.1} has
pointers $x$, $y$, and $z$ all pointing to the same buffer $b_1$.

\subsection{Modelling Buffers and Buffer Pointer Relations}
\label{sec:model}

We identify a buffer by its allocation site name and record its size
and the set of positions in the buffer that hold the null character
$'\backslash 0'$.

Let \nds be the set of nodes in the control flow graph of the program being analysed, \int be the set of positive integers 
(including 0) and \pts be the set of pointers to
buffers. Using these sets, we define \text{$	\biinfL  = \{ b_n \mid n \in \nds \} \cup \{b_\infty\}$} as a set of buffer identities, 
	\text{$	\sizeL  = \int \cup \{\infty\}$} as a set of buffer sizes (or offsets into buffers), and 
	\text{$	\nullposL  = 2^{\sizeL}$} as the set of positions of null characters in a buffer. 
Buffers are described using the following functions.

\begin{itemize}
\item \text{$\buf = \biinfL \mapsto \sizeL \times \nullposL$} maps a buffer identity to buffer information.

       A buffer is identified by the program point of its creation {\rm e.g.}
       $b_n$ denotes the buffer created in statement \text{$n\in\nds$}. 
       Each buffer has exactly one size and exactly one set of null
       positions at any given program point.  \buf also contains a fictitious 
	``undefined'' buffer \text{$(b_\infty, \infty, \emptyset)$}. 

       Buffer size $\infty$ may also be associated with a valid buffer indicating that 
       at that program point, the buffer size
       is not a compile time constant. This could be either because
       the buffer has not been created along some execution path, or
       the buffer has been created using a size that is not known at
       compile time, or the memory has been freed along some execution
       path.\footnote{Since a buffer is identified by its program point
       of creation, known but dissimilar buffer sizes along different
       execution paths reaching a program point are not possible.} 

       If the null position set is empty, it indicates that a read will
       cause a buffer overflow. The presence of $\infty$ in a null
       position set indicates that a write has already caused buffer
       overflow.

\item \text{$\bpt = \pts \mapsto \left(\biinfL \times \sizeL\right)$} relates a buffer pointer
      \text{$x \in \pts$} to its pointee buffer
      \text{$b_n \in \biinfL$} and an offset \text{$i
      \in \sizeL$} into the buffer (because a pointer may point to some
      position in the middle of a buffer). 

\end{itemize}

\begin{figure}[!t]
\begin{tabular}{@{}c@{}|@{}c@{}}
\begin{tabular}{@{}c@{}}
\begin{minipage}{55mm}
\begin{verbatim}
/* Initial situation:
   w points to buffer b0
   x points to buffer b1*/

    z = x+6;
/* After this, z points 
    to offset 6 in b1   */

    y = x+4;
/* After this, y points 
    to offset 4 in b1   */
  
/* Call 1. No overflow  */
    strcat(z,y); 

/* Call 2. b1 overflows */
    strcat(z,y); 

/* Call 3. No overflow  */
    strcpy(y,w); 
\end{verbatim}
\end{minipage}
\\\\
(a) Program
\end{tabular}
&
\begin{tabular}{@{}c@{}}
\begin{pspicture}(0,-2)(110,38)
\putnode[l]{o1}{origin}{22}{23}{}
\putnode[l]{f1}{o1}{-4}{-6}{\raisebox{2.25mm}{\psframebox{\rule{89mm}{0mm}\rule{0mm}{4.2mm}}}}
\putnode[l]{m1}{o1}{-2}{-5}{\rule{0mm}{3mm}\tt o}
\putnode[l]{m2}{m1}{6}{0}{\rule{0mm}{3mm}\tt n}
\putnode[l]{m3}{m2}{6}{0}{\rule{0mm}{3mm}\tt e}
\putnode[l]{m4}{m3}{6}{0}{\rule{0mm}{3mm}\!\!\!\raisebox{-.95mm}{\tt'\!\char`\\0\!'}}
\putnode[l]{m5}{m4}{6}{0}{\rule{0mm}{3mm}\tt t}
\putnode[l]{m6}{m5}{6}{0}{\rule{0mm}{3mm}\tt w}
\putnode[l]{m7}{m6}{6}{0}{\rule{0mm}{3mm}\tt o}
\putnode[l]{m8}{m7}{6}{0}{\rule{0mm}{3mm}\!\!\!\raisebox{-.95mm}{\tt'\!\char`\\0\!'}}
\putnode[l]{m9}{m8}{6}{0}{\rule{0mm}{3mm}\tt t}
\putnode[l]{m10}{m9}{6}{0}{\rule{0mm}{3mm}\tt h}
\putnode[l]{m11}{m10}{6}{0}{\rule{0mm}{3mm}\tt r}
\putnode[l]{m12}{m11}{6}{0}{\rule{0mm}{3mm}\tt e}
\putnode[l]{m13}{m12}{6}{0}{\rule{0mm}{3mm}\tt e}
\putnode[l]{m14}{m13}{6}{0}{\rule{0mm}{3mm}\!\!\!\raisebox{-.95mm}{\tt'\!\char`\\0\!'}}
\putnode[l]{m15}{m14}{6}{0}{??}
\putnode[l]{n1}{m1}{0}{7}{\footnotesize 0}
\putnode[l]{n2}{m2}{0}{7}{\footnotesize 1}
\putnode[l]{n3}{m3}{0}{7}{\footnotesize 2}
\putnode[l]{n4}{m4}{0}{7}{\footnotesize 3}
\putnode[l]{n5}{m5}{0}{7}{\footnotesize 4}
\putnode[l]{n6}{m6}{0}{7}{\footnotesize 5}
\putnode[l]{n7}{m7}{0}{7}{\footnotesize 6}
\putnode[l]{n8}{m8}{0}{7}{\footnotesize 7}
\putnode[l]{n9}{m9}{0}{7}{\footnotesize 8}
\putnode[l]{n10}{m10}{0}{7}{\footnotesize 9}
\putnode[l]{n11}{m11}{0}{7}{\footnotesize 10}
\putnode[l]{n12}{m12}{0}{7}{\footnotesize 11}
\putnode[l]{n13}{m13}{0}{7}{\footnotesize 12}
\putnode[l]{n14}{m14}{0}{7}{\footnotesize 13}
\putnode[l]{n15}{m15}{0}{7}{\footnotesize 14}

\psset{arrowsize=2}
\putnode[l]{a1}{n1}{10}{-20}{\psframebox{\rule{2mm}{0mm}\raisebox{1mm}{\rnode{x1}{}}\rule{0mm}{2mm}\rule{2mm}{0mm}}}
\putnode[l]{u}{a1}{2}{-5}{\,x}
\putnode[l]{a2}{a1}{12}{0}{\psframebox{\rule{2mm}{0mm}\raisebox{1mm}{\rnode{y1}{}}\rule{0mm}{2mm}\rule{2mm}{0mm}}}
\putnode[l]{u}{a2}{2}{-5}{\,y}
\putnode[l]{a3}{a2}{12}{0}{\psframebox{\rule{2mm}{0mm}\raisebox{1mm}{\rnode{z1}{}}\rule{0mm}{2mm}\rule{2mm}{0mm}}}
\putnode[l]{u}{a3}{2}{-5}{\,z}
\putnode[l]{a4}{a1}{-12}{0}{\psframebox{\rule{2mm}{0mm}\raisebox{1mm}{\rnode{w1}{}}\rule{0mm}{2mm}\rule{2mm}{0mm}}}
\putnode[l]{u}{a4}{2}{-5}{w}
\nccurve[ncurv=.8,angleA=90,angleB=300,offsetB=1.5,nodesepB=1.5]{*->}{x1}{m1}
\nccurve[ncurv=.5,angle=90,angleB=270,nodesepB=2.25]{*->}{y1}{m5}
\nccurve[ncurv=.45,angle=90,angleB=270,nodesepB=2.25]{*->}{z1}{m7}
\putnode[l]{p}{o1}{0}{15}{}
\putnode[l]{f2}{p}{-4}{-6}{\raisebox{2.25mm}{\psframebox{\rule{65mm}{0mm}\rule{0mm}{4.2mm}}}}
\putnode[l]{m1}{p}{-2}{-5}{\rule{0mm}{2mm}\tt n}
\putnode[l]{m2}{m1}{6}{0}{\rule{0mm}{2mm}\tt o}
\putnode[l]{m3}{m2}{6}{0}{\rule{0mm}{2mm}}
\putnode[l]{m4}{m3}{6}{0}{\rule{0mm}{2mm}\tt p}
\putnode[l]{m5}{m4}{6}{0}{\rule{0mm}{2mm}\tt r}
\putnode[l]{m6}{m5}{6}{0}{\rule{0mm}{2mm}\tt o}
\putnode[l]{m7}{m6}{6}{0}{\rule{0mm}{2mm}\tt b}
\putnode[l]{m8}{m7}{6}{0}{\rule{0mm}{2mm}\tt l}
\putnode[l]{m9}{m8}{6}{0}{\rule{0mm}{2mm}\tt e}
\putnode[l]{m10}{m9}{6}{0}{\rule{0mm}{2mm}\tt m}
\putnode[l]{m11}{m10}{6}{0}{\rule{0mm}{2mm}\!\!\!\raisebox{-.95mm}{\tt'\!\char`\\0\!'}}
\nccurve[angle=150,angleB=240,offsetB=-1.25,nodesepB=2.25]{*->}{w1}{m1}
\putnode[r]{w}{f1}{0}{0}{%
	\begin{tabular}{l}
	buffer $b_1$
	\\
	size 14
	\end{tabular}
	}
\putnode[r]{w}{f2}{0}{0}{%
	\begin{tabular}{l}
	buffer $b_0$
	\\
	size 10
	\end{tabular}
	}

\end{pspicture}
\\
(b) Memory before call 1. ``??'' indicates garbage value.
\\
\\ \hline
\\
\begin{pspicture}(0,-2)(110,38)
\putnode[l]{o1}{origin}{22}{23}{}
\putnode[l]{f1}{o1}{-4}{-6}{\raisebox{2.25mm}{\psframebox{\rule{89mm}{0mm}\rule{0mm}{4.2mm}}}}
\putnode[l]{m1}{o1}{-2}{-5}{\rule{0mm}{3mm}\tt o}
\putnode[l]{m2}{m1}{6}{0}{\rule{0mm}{3mm}\tt n}
\putnode[l]{m3}{m2}{6}{0}{\rule{0mm}{3mm}\tt e}
\putnode[l]{m4}{m3}{6}{0}{\rule{0mm}{3mm}\!\!\!\raisebox{-.95mm}{\tt'\!\char`\\0\!'}}
\putnode[l]{m5}{m4}{6}{0}{\rule{0mm}{3mm}\tt t}
\putnode[l]{m6}{m5}{6}{0}{\rule{0mm}{3mm}\tt w}
\putnode[l]{m7}{m6}{6}{0}{\rule{0mm}{3mm}\tt o}
\putnode[l]{m8}{m7}{6}{0}{\rule{0mm}{3mm}\tt t}
\putnode[l]{m9}{m8}{6}{0}{\rule{0mm}{3mm}\tt w}
\putnode[l]{m10}{m9}{6}{0}{\rule{0mm}{3mm}\tt o}
\putnode[l]{m11}{m10}{6}{0}{\rule{0mm}{3mm}\!\!\!\raisebox{-.95mm}{\tt'\!\char`\\0\!'}}
\putnode[l]{m12}{m11}{6}{0}{\rule{0mm}{3mm}\tt e}
\putnode[l]{m13}{m12}{6}{0}{\rule{0mm}{3mm}\tt e}
\putnode[l]{m14}{m13}{6}{0}{\rule{0mm}{3mm}\!\!\!\raisebox{-.95mm}{\tt'\!\char`\\0\!'}}
\putnode[l]{m15}{m14}{6}{0}{??}
\putnode[l]{n1}{m1}{0}{7}{\footnotesize 0}
\putnode[l]{n2}{m2}{0}{7}{\footnotesize 1}
\putnode[l]{n3}{m3}{0}{7}{\footnotesize 2}
\putnode[l]{n4}{m4}{0}{7}{\footnotesize 3}
\putnode[l]{n5}{m5}{0}{7}{\footnotesize 4}
\putnode[l]{n6}{m6}{0}{7}{\footnotesize 5}
\putnode[l]{n7}{m7}{0}{7}{\footnotesize 6}
\putnode[l]{n8}{m8}{0}{7}{\footnotesize 7}
\putnode[l]{n9}{m9}{0}{7}{\footnotesize 8}
\putnode[l]{n10}{m10}{0}{7}{\footnotesize 9}
\putnode[l]{n11}{m11}{0}{7}{\footnotesize 10}
\putnode[l]{n12}{m12}{0}{7}{\footnotesize 11}
\putnode[l]{n13}{m13}{0}{7}{\footnotesize 12}
\putnode[l]{n14}{m14}{0}{7}{\footnotesize 13}
\putnode[l]{n15}{m15}{0}{7}{\footnotesize 14}

\psset{arrowsize=2}
\putnode[l]{a1}{n1}{10}{-20}{\psframebox{\rule{2mm}{0mm}\raisebox{1mm}{\rnode{x1}{}}\rule{0mm}{2mm}\rule{2mm}{0mm}}}
\putnode[l]{u}{a1}{2}{-5}{\,x}
\putnode[l]{a2}{a1}{12}{0}{\psframebox{\rule{2mm}{0mm}\raisebox{1mm}{\rnode{y1}{}}\rule{0mm}{2mm}\rule{2mm}{0mm}}}
\putnode[l]{u}{a2}{2}{-5}{\,y}
\putnode[l]{a3}{a2}{12}{0}{\psframebox{\rule{2mm}{0mm}\raisebox{1mm}{\rnode{z1}{}}\rule{0mm}{2mm}\rule{2mm}{0mm}}}
\putnode[l]{u}{a3}{2}{-5}{\,z}
\putnode[l]{a4}{a1}{-12}{0}{\psframebox{\rule{2mm}{0mm}\raisebox{1mm}{\rnode{w1}{}}\rule{0mm}{2mm}\rule{2mm}{0mm}}}
\putnode[l]{u}{a4}{2}{-5}{w}
\nccurve[ncurv=.8,angleA=90,angleB=300,offsetB=1.5,nodesepB=1.5]{*->}{x1}{m1}
\nccurve[ncurv=.5,angle=90,angleB=270,nodesepB=2.25]{*->}{y1}{m5}
\nccurve[ncurv=.45,angle=90,angleB=270,nodesepB=2.25]{*->}{z1}{m7}
\putnode[l]{p}{o1}{0}{15}{}
\putnode[l]{f2}{p}{-4}{-6}{\raisebox{2.25mm}{\psframebox{\rule{65mm}{0mm}\rule{0mm}{4.2mm}}}}
\putnode[l]{m1}{p}{-2}{-5}{\rule{0mm}{2mm}\tt n}
\putnode[l]{m2}{m1}{6}{0}{\rule{0mm}{2mm}\tt o}
\putnode[l]{m3}{m2}{6}{0}{\rule{0mm}{2mm}}
\putnode[l]{m4}{m3}{6}{0}{\rule{0mm}{2mm}\tt p}
\putnode[l]{m5}{m4}{6}{0}{\rule{0mm}{2mm}\tt r}
\putnode[l]{m6}{m5}{6}{0}{\rule{0mm}{2mm}\tt o}
\putnode[l]{m7}{m6}{6}{0}{\rule{0mm}{2mm}\tt b}
\putnode[l]{m8}{m7}{6}{0}{\rule{0mm}{2mm}\tt l}
\putnode[l]{m9}{m8}{6}{0}{\rule{0mm}{2mm}\tt e}
\putnode[l]{m10}{m9}{6}{0}{\rule{0mm}{2mm}\tt m}
\putnode[l]{m11}{m10}{6}{0}{\rule{0mm}{2mm}\!\!\!\raisebox{-.95mm}{\tt'\!\char`\\0\!'}}
\nccurve[angle=150,angleB=240,offsetB=-1.25,nodesepB=2.25]{*->}{w1}{m1}
\putnode[r]{w}{f1}{0}{0}{%
	\begin{tabular}{l}
	buffer $b_1$
	\\
	size 14
	\end{tabular}
	}
\putnode[r]{w}{f2}{0}{0}{%
	\begin{tabular}{l}
	buffer $b_0$
	\\
	size 10
	\end{tabular}
	}

\end{pspicture}
\\
(c) Memory before call 2. ``??'' indicates garbage value.
\\
\end{tabular}
\\\hline
\multicolumn{2}{c}{}
\\
\multicolumn{2}{@{}c@{}}{%
\begin{tabular}{@{}c@{}}
\begin{minipage}{150mm}
\end{minipage}
\[
\begin{array}{@{}l|l|l}
\hline
\text{Relevant program points}
	& \begin{tabular}{@{}l@{}}Buffer and pointer \\ mappings\end{tabular}
	& \text{Relevant extractor functions}
	\\ \hline\hline
\text{Before call 1}
&
\begin{array}{@{}l@{}l@{}}
\bm  = \{ \begin{array}[t]{@{}l@{}}
		(b_0, 10, \{10\}),
		\\
		(b_1, 14, \{3, 7, 13\})
		\}
	   \end{array}
\\
\bp = \{  \begin{array}[t]{@{}l@{}}
		(w, b_0, 0), 
		(x, b_1, 0), 
		\\
		(y, b_1, 4), 
	 (z, b_1, 6) \}
	\end{array}
\end{array}
&
\begin{array}{@{}l@{}}
	\size(\bm, b_0) = 10,
	\nps(\bm, b_0) = \{10\}
\\
	\size(\bm, b_1) = 14,
	\nps(\bm, b_1) = \{3, 7, 13\}
\\ \hline
	\pointee(\bp, w) =  b_0 , \;\; start(\bp,w,b_0) = 0
\\
	\pointee(\bp, x) =  b_1 , \;\; start(\bp,x,b_1) = 0
\\
	\pointee(\bp, y) =  b_1 , \;\; start(\bp,y,b_1) = 4
\\
	\pointee(\bp, z) =  b_1 , \;\; start(\bp,z,b_1) = 6
\end{array}
\\ \hline
\renewcommand{\arraystretch}{.9}%
\begin{tabular}[t]{@{}l@{}}
After call 1, null positions in $b_1$ \\ change. No other change.
\end{tabular}
&
\bm  = \{ \begin{array}[t]{@{}l@{}}
		(b_0, 10, \{10\}),
		\\
		(b_1, 14, \{3, 10, 13\})
		\}
	   \end{array}
&
\begin{array}[t]{@{}l@{}}
	\size(\bm, b_1) = 14
\\
	\nps(\bm, b_1) = \{3, 10, 13\}
\end{array}
\\ \hline
\renewcommand{\arraystretch}{.9}%
\begin{tabular}[t]{@{}l@{}}
After call 2, null positions in $b_1$ \\ change again. No other change.
\end{tabular}
&
\bm  = \{ \begin{array}[t]{@{}l@{}}
		(b_0, 10, \{10\}),
		\\
		(b_1, 14, \{3, \infty\})
		\}
	   \end{array}
&
\begin{array}[t]{@{}l@{}}
	\size(\bm, b_1) = 14
\\
	\nps(\bm, b_1) = \{3, \infty\}
\end{array}
\\ \hline
\renewcommand{\arraystretch}{.9}%
\begin{tabular}[t]{@{}l}
After call 3, null positions in $b_1$ \\ change again. No other change.
\end{tabular}
&
\bm  = \{ \begin{array}[t]{@{}l@{}}
		(b_0, 10, \{10\}),
		\\
		(b_1, 14, \{3, 14, \infty\})
		\}
	   \end{array}
&
\begin{array}[t]{@{}l@{}}
	\size(\bm, b_1) = 14
\\
	\nps(\bm, b_1) = \{3, 14, \infty\}
\end{array}
\\ \hline
\end{array}
\]
\\ \\
(d) Modelling buffers and buffer pointers. Presence of $\infty$ indicates that an overflow has occurred some time.
\end{tabular}
}
\\ \hline
\end{tabular}

\caption{Example to illustrate modelling of buffers and buffer pointers.}
\label{fig:exmp.1}
\end{figure}
We assume that both these functions are total functions. This simplifies our formulations.

We use the following notational conventions: 
\begin{itemize}[label=$-$]
\item \text{$\bm$} ranges over the set $A=2^\buf$ (and thus represent buffer mappings).
\item \text{$\bp$} ranges over the set $B=2^\bpt$ (and thus represent pointer to buffer mappings). 
\item \text{$b$} ranges over buffer identities in $\biinfL$. 
\item \text{$w, x,y,z$} range over the set of pointers $\pts$.
\item \text{$i,j,k, l, m$} range over the set  \sizeL.
\item \text{$X,Y,Z$} range over some set (the types of their elements are evident from the context).
\end{itemize}
For a buffer map $\bm$, the extractor functions {\sf\em
size$(\bm,b)$\/} and \text{$\nps(\bm,b)$} compute the size and the
null position set of a given buffer $b$ in \bm.
\begin{align}
(b,k,X) \in \bm \Rightarrow \size{(\bm, b)} =k \wedge \nps(\bm, b) =X
	\label{eq:fun.size}
\end{align}
In other words, \text{$\bm(b) = (\size{(\bm,b)}, \nps(\bm,b))$}.  

For a buffer pointer map $\bp$, the extractor functions \text{$\pointee(\bp,x)$} and 
       \text{$\start(\bp,x,b)$} return the pointee buffer and the start position of $x$ in a given buffer $b$.
\begin{align}
(x,b,i) \in \bp \Rightarrow \pointee(\bp, x) = b \wedge \start(\bp, x, b) =i
	\label{eq:fun.start}
\end{align}
We compute the positions of null characters and start positions using
compile time saturated addition of integers that limits the result to
the given (compile time) constant $k$ as defined below.
\begin{align}
\forall\, i, j \in \nullposL\!:\;\;
i \oplus_k j &=
	\begin{cases}
	i + j & i + j  \leq k \wedge k \neq \infty
		\\
	\infty	& \text{otherwise}
	\end{cases}
\end{align}

\subsubsection*{Running Example}

Figure~\ref{fig:exmp.1} illustrates our modelling. 
The details of these computations are explained in Section~\ref{sec:exmp}.

\subsection{Data Flow Equations}

This section describes the lattices of data flow values, the data flow equations, and the flow functions used in the data flow equations.

\subsubsection{Lattices}

Our analysis computes subsets of \buf and \bpt flow sensitively at each
node \text{$n \in \nds$}. The lattices of these values are \text{$(
A, \aO)$} and \text{$(B, \bO\rangle$} where
\text{$A=2^\buf$} and \text{$B=2^\bpt$}. The overall lattice $L$ is the
product lattice \text{$(A\times B, \abO)$}.
Its meet operation \abM is defined in terms of the meet operations
\aM and $\bM$ of the constituent lattices $A$ and $B$. 
\begin{align}
\forall \, \bm,\bm' \in A, \;
\forall \, \bp,\bp' \in B\!: \;\;
(\bm,\bp) \,\abM (\bm',\bp') 
	&= (\bm \,\aM \bm', \bp \,\bM \bp') 
\end{align}

We first define the meet \aM for merging the buffers (equations~\ref{eq:meet.A} to~\ref{eq:meet.NP})
and then define the meet \bM for merging the pointer buffer mappings (equation~\ref{eq:meet.B}).

\subsubsection*{Merging Buffers}

Recall that lattice \text{$(A=2^\buf,\aO)$} 
where \text{$\buf = \biinfL \mapsto \sizeL \times \nullposL$} is defined in terms of lattices
\text{$(\sizeL,\sizeO)$} and 
\text{$(\nullposL,\nullposO)$}.  Hence \aM is defined in terms
of \sizeM and \nullposM.
\begin{align}
\forall \, \bm,\bm' \in A\!: \;\;
\bm \,\aM \bm' &= \left\{ \left(b, \; 
				\left(
		\left(i \,\sizeM j , \; X \,\nullposM Y \right) 
				\right)
				\right) 
				\; \middle| \; \left(b,i, X\right) \in \bm, \left(b,j, Y\right) \in \bm' 
			\right\}	
	\label{eq:meet.A}
\end{align}
The definitions of
\sizeM and \nullposM warrant some explanation.

\begin{itemize}
\item The meet \sizeM enforces a buffer to have the same size along
      all paths. In case of inconsistent sizes, the buffer size is
      recorded as $\infty$ which indicates that the buffer size is statically
      undefined (see Section~\ref{sec:model} for the semantics of $\infty$).
      \begin{align}
      \forall \, i,j \in \sizeL\!: \;\; i\, \sizeM j 
		      &=
			      \begin{cases}
			      i  & i = j 
				      \\
			      \infty & \text{otherwise}
			      \end{cases}
	\label{eq:meet.SZ}
      \end{align}
      The definition of \sizeM indicates that $\infty$ is the $\bot$ element of
      \text{$(\sizeL,\sizeO)$}. However, it does not contain a $\top$ and hence
      is only a meet semilattice. 

\item 
      Given the soundness requirement described in
      Section~\ref{sec:assumptions}, the \nullposM should approximate the null
      position sets of buffers by intersecting them. This effectively
      lengthens the strings in the buffer guaranteeing the soundness of
      this approximation. However, this idea has a minor flaw: Assume that for
      a given buffer, we get the set \text{$\{1,5,14\}$} along one path
      and \text{$\{1,7, \infty\}$} along the other path. We would like
      to assume that we have a null character at position 1 alone and
      intersection indeed achieves this. However, we cannot ignore the
      buffer overflow that has occurred along the second control flow
      path. An intersection would remove $\infty$ from the resulting
      set because it appears in only one set. 

      We overcome this problem by observing that a set of null positions
      $X$ is a subset of \text{$\sizeL = \int \cup \{\infty\}$}. Hence we
      define two function $\fin(X)$ and $\inf(X)$ that 
      partition $X$ into subsets of \int and \text{$\{\infty\}$}: $\fin(X)$ computes 
      the maximal subset of $X$ containing finite numbers, and $\inf(X)$ computes the minimal
      subset of $X$ containing $\infty$.

      Consider the following sets \text{$X_1 = \{2, 5, 10, \infty\}$},
      \text{$X_2 = \{2, 5, 10\}$}, \text{$X_3 = \{\infty\}$}, and
      \text{$X_4 = \emptyset$}. Then,
	\begin{itemize}
      \item \text{$\fin(X_1) = \{2, 5, 10\}$}, and \text{$\inf(X_1) = \{ \infty\}$}. 
      \item \text{$\fin(X_2) = \{2, 5, 10\}$}, and \text{$\inf(X_2) = \emptyset$}. 
      \item \text{$\fin(X_3) = \emptyset$}, and \text{$\inf(X_3) = \{ \infty\}$}.
      \item \text{$\fin(X_4) =  \inf(X_4) = \emptyset$}.
	\end{itemize}
      This distinction allows us to use intersection for $\fin(X)$ and union for $\inf(X)$ in the definition of
      \nullposM. 
      \begin{align}
	\forall \, X,Y \subseteq \nullposL\!: \;\;
      X \,\nullposM Y 
		      &=
			      \left(\fin(X) \cap \fin(Y)\right)\cup \inf(X) \cup \inf(Y)
	\label{eq:meet.NP}
      \end{align}
\end{itemize}

\subsubsection*{Merging Pointer Buffer Mappings}

Buffer mappings \text{$\bp, \bp' \in B$} are merged using \bM.
Since \text{$B=2^\bpt$} where
\text{$\bpt = \pts \mapsto \left(\biinfL \times \sizeL\right)$},
\bM is defined in terms of 
\bidM and \sizeM: 
\begin{align}
\forall \, \bp,\bp' \in B\!: \;\;
\bp \,\bM \bp'
	& = 
	   \left\{
		\left( x, b \,\bidM b', k \,\sizeM k' \right)
		\;\middle| \;
		\left( x, b, k \right) \in \bp, 
		\left( x, b', k' \right) \in \bp'  
	   \right\}
	\label{eq:meet.B}
\end{align}
The meet operation \sizeM is defined in equation~(\ref{eq:meet.SZ}) earlier; 
\bidM is similarly defined below.
      \begin{align}
\forall \, b,b' \in \biinfL\!: \;\;
      b \,\bidM b' 
		      &=
			      \begin{cases}
			      b  & b = b' 
				      \\
			      b_\infty & \text{otherwise}
			      \end{cases}
      \end{align}
\bidM imposes the restriction that a pointer points to the
same buffer along all paths reaching a program point; \sizeM ensures that the offsets are also identical.

Since \biinfL and \sizeL are meet semilattice, their product
\text{$\biinfL\times\sizeL$} is also a meet semilattice. 
It does not have a $\top$ element and
its $\bot$ element is \text{$(b_\infty,\infty)$}. 

\begin{figure}[!t]
\[
\begin{array}{|l|c|c|l@{}|}
\hline
\multicolumn{4}{|l|}{%
	\begin{tabular}{@{}l@{}}
		For brevity, we use the following short forms
			\\
		$\bullet$ \; Destination buffer: $D_n$, New size: $K_n(\bm,\bp)$, New relevant null position set: $R_n(\bm,\bp)$
			\\
		$\bullet$ \; $\pointee(\bp,z) = pt_z, \size(\bm,\pointee_z) = \size{}_z, \nps(\pointee_z) = \nps_z$,
		$ \start(\bp,z,\pointee_z) = \start_z, \strend(\bm,\bp,z,\pointee_z) = \strend_z $
			\\
	\end{tabular}
		}
	\\ \hline
\text{Statement } n 
	& D_n
	& K_n(\bm,\bp)
	& R_n(\bm,\bp)
	\\ \hline\hline
x = \malloc(m)
	& b_n
	& m
	& \emptyset
	\\\hline
x = \calloc(m)
	& b_n
	& m
	& \{ i \mid i \leq m \}
	\\\hline
\free(x)
	& \pointee_x
	& 0
	& \nps_x
	\\\hline
x[i] = '\backslash 0'
	& \pointee_x
	& \size{}_x
	& \nps_x \cup \{ i \oplus_{\size{}_x} \start_x \}
	\\\hline
x[i] = c
	& \pointee_x
	& \size{}_x
	& \nps_x - \{ i \oplus_{\size{}_x} \start_x \}
	\\\hline
\multirow{7}{*}{%
\begin{tabular}{@{}l@{}}
\strcpy$(x,y)$
	\\
\strcat$(x,y)$
	\\
\strncpy$(x,y,m)$
	\\
\strncat$(x,y,m)$
\end{tabular}
}%
	& 
\multirow{7}{*}{$
		\pointee_x
$}
	& 
\multirow{7}{*}{$
	\size{}_x
$}
	& 
	  \begin{array}{@{}ll@{}}
	\text{\em let}
	\\
	& \begin{array}{@{}r@{\ }l}
		\nullpos_y & = \begin{cases}
				\strend_y - \start{}_y
					& \strcpy \text{ or } \strcat
				\\
				\min(m\!+\!1, \strend_y - \start{}_y)
					& \strncpy \text{ or } \strncat
				\end{cases}
			\\
		\rule{0em}{2.em}%
		\copypos_x & = \begin{cases}
				\start{}_x \hspace*{30mm}
					& \strcpy \text{ or } \strncpy
				\\
				\strend_x 
					& \strcat \text{ or } \strncat
				\end{cases}
			\\
		\npfromy_{xy} & = \nullpos_y \oplus_{\size{}_x} \copypos_x 
			\\
	    	\npsbefore_{xy} & = \rnps(\nps_x,\size{}_x, 0, \start_x, <  ) 
			\\
		\npat_{xy}   &=  \{\npfromy_{xy}\} 
			\\
	    	\npsafter_{xy} &= \rnps(\nps_x,\size{}_x, 0, \npfromy_{xy}, \geq  ) 
	  \end{array}
	\\
	\text{\em in}
	\\
	   & \npsbefore_{xy} \; \cup \; \npat_{xy} \;\cup \; \npsafter_{xy}
	  \end{array}
	\\\hline
\memcpy(x,y,m)
	& \pointee_x
	& \size{}_x
	& 
	  \begin{array}{@{}ll@{}}
	\text{\em let}
	\\
	& 
		\begin{array}{r@{\ }l}
		\sdist_{xy} & = \start_x-\start_y 
				\\
		\npsfromy_{xy}& = \rnps(\nps_y,\size{}_x,\sdist_{xy}, \start_y, \geq ) \cap
			\\
		      & \hspace*{4mm}\rnps(\nps_y,\size{}_x,\sdist_{xy}, \start_y + m, \leq )
			\\
	    	\npsbefore_{xy} & = \rnps(\nps_x,\size{}_x, 0, \start_x, <  ) 
			\\
	    	\npsafter_{xy} &= \rnps(\nps_x,\size{}_x, 0, \start_x + m, \geq  ) 
			\\
		\oflow_{xy}  & = \{m\oplus_{\size{}_x}\start_x \}  - \{m+\start_x\} \; \cup
			\\
		 	& \hspace*{4mm} \{m\oplus_{\size{}_y}\start_y \}  - \{m+\start_y\} 
		\end{array}
	\\
	in
	\\	
	&  \begin{array}{@{}l@{}}
	    	\npsbefore_{xy}  \; \cup \npsafter_{xy}  \; \cup \; \npsfromy_{xy}
			\; \cup \; \oflow_{xy} 
	  \end{array}
	\end{array}
	\\ \hline
\end{array}
\]
\caption{Buffer mapping extractor functions for statements. We assume that $m$ and $i$ 
         are compile time constants or appropriate limits of ranges (see Section~\ref{sec:assumptions}).}
\label{fig:flow.fun}
\end{figure}

\subsubsection{Data Flow Equations}

Recall that our data flow values are pairs \text{$(\bm,\bp)$} that record the buffer mappings and pointer 
to buffer mappings at each program point.

The data flow equations are as follows:
\begin{align}
\bin_n & = 
	\begin{cases}
	\left(\left\{ \left(b_n, \infty, \emptyset\right) \;\middle|\; n \in \nds \right\}, 
	\left\{ \left(x, b_\infty, \infty \right) \;\middle|\; x \in \pts\right\}
        \right)
		& n = \Start{}
		\\
	\displaystyle{\bigsqcap_{p \in pred(n)}}{\hspace*{-4mm}\raisebox{-2mm}{$\scriptstyle AB$}}
	\bout_p
		& \text{otherwise}
	\end{cases}
	\\
\bout_n & = \updatemaps(\bin_n,n)
\end{align}
At the start of the program, all buffers are undefined (i.e. their sizes are $\infty$ and their
null pointer sets are $\emptyset$) and all
pointers point to undefined buffer \text{$(b_\infty,\infty)$}. We need this boundary information instead 
of empty mappings $\emptyset$ because we need to maintain the invariant that our mappings are total functions.

Flow function \updatemaps updates the buffer mappings and pointer to buffer mappings.
\begin{align}
\updatemaps(\bm,\bp,n)
	& = 
		(
		\updatebuf(\bm,\bp,n),
		\updatebpt(\bm,\bp,n)
		)
\end{align}
For statement $n$, the destination buffer $D_n$ in buffer mapping \bm is updated using auxiliary 
extractor functions $K_n(\bm,\bp)$ and $R_n(\bm,\bp)$; 
the three terms are defined for relevant statements Figure~\ref{fig:flow.fun} and are explained
Section~\ref{sec:flow.func}.
\begin{align}
\updatebuf(\bm,\bp,n)
	&= \begin{cases}
	    \bm\left[D_n \mapsto \left(
			K_n(\bm,\bp),
			R_n(\bm,\bp)\right)\right]
			& \renewcommand{\arraystretch}{.9}
			\begin{array}{@{}l@{}}
			\text{Statement $n$ involves a string } 
				\\
			   \text{function listed in Figure \ref{fig:flow.fun}}
			\end{array}
			\\
			\bm & \text{Otherwise}
	    \end{cases}
\end{align}
The pointer to buffer maps are updated only for the statements which 
involve a pointer assignment. 
\begin{align}
\updatebpt(\bm,\bp,n)
		&=
		\begin{cases}
		\bp[x \mapsto \{ (b_n,0) \}]
			& \renewcommand{\arraystretch}{.9}
			\begin{array}{@{}l@{}}
			\text{Statement $n$ is } x = \malloc(k) \text{ or } x = \calloc(k) 
				\\
			   \text{or } x = \str(k)
			\end{array}
			\\
		\bp[x \mapsto \bp(y)]
			& \text{Statement $n$ is } x = y
			\\
	 	\bp[x \mapsto \adv(\bm,\bp,y,i)]
			& \text{Statement $n$ is } x = y + i
			\\
		\bp
			& \text{Otherwise}
		\end{cases}
\end{align}
where \text{$\adv(\bm,\bp,y,i)$} is used to compute the 
offset of $y$ in the buffer by shifting it by $i$.
\begin{align}
\adv(\bm,\bp,y,i) &=
	\{ (b, j \oplus_k i) \mid b = \pointee(\bp,y), j = \start(\bp,y,b), k = \size{(\bm,b)})\} 
\end{align}

\subsubsection{Flow Functions for Computing the Set of Null Positions}
\label{sec:flow.func}

The heart of this analysis lies in computing the set of positions of null characters. We use the following 
functions to achieve this:
\begin{enumerate}
\item Functions \size{}, \nps, \pointee, and \start introduced in equations (\ref{eq:fun.size}) and 
      (\ref{eq:fun.start}).
\item Function \rnps which performs arithmetic on null positions to compute the relevant null positions. 
      \begin{align}
      \rnps(X, \saturate, \shift, \pivot,\relop )
	      &= 
		      \{ i \oplus_\saturate \shift \mid i \in X, i \;\relop \,\pivot \} 
      \end{align}
      It is defined in terms of
      \begin{itemize}
      \item a set of null positions ($X$), 
      \item a saturation limit (\saturate), 
      \item an offset (\shift) in the destination buffer using which the null positions of the source buffer may have to be shifted, 
      \item null positions in the source buffer described in terms of
      \begin{itemize}
      \item a pivot (\pivot) around which the null positions are to be examined, and 
      \item a relational operator (\text{$\relop \in \{ <, \leq, >, \geq, =, \neq \}$}) 
            used for comparison with the pivot.
      \end{itemize}
      \end{itemize}
\item Function \strend which computes the end of a string in a buffer using \rnps by finding out
      the null positions that lie beyond the \start of the string and then computing their
      greatest lower bound (\glb) under $\leq$ as the partial order.
      This allows us to restrict the \rnps set in a buffer to the first null position after the start
      position of a given string.
      \begin{align}
      \strend(\bm,\bp,x, b)
	      &= 
		      \glb(\rnps(\nps(b),\size(\bm, b),0,\start(\bp,x,b), \geq))
      \end{align}
      Although our primary interest is in 
      computing the minimum among the null positions, \glb is more convenient. It inherently models
      the situation when no null position is found because $\glb(\emptyset) = \infty$ indicating that the string is
      infinitely long (i.e. the string overflows the buffer).\footnote{%
      Given that
      \glb computes the greatest number that is smaller than any number in the given set, if the given set does not contain 
      any number, the greatest number that is smaller than no number is $\infty$.} 

\end{enumerate}

\subsubsection*{Set of Null Positions for String Functions}

We use the above functions by dividing the null position sets resulting from a string operation into three categories 
based on the start position $s$ and the end position $e$ of the copied string in the destination buffer. Given
$x$ as the destination pointer and $y$ as the source pointer:
\begin{enumerate}
\item \npsbefore. The null positions that remain unchanged in the destination buffer because they appear before $s$.
	This is defined by \text{$\rnps(\nps_x,\size{}_x, 0, \start_x, <  ) $} in Figure~\ref{fig:flow.fun}.

\item \npat. The null position appearing at $e$. 
	This  is defined by \npfromy in Figure~\ref{fig:flow.fun}. It
		represents the null position imported from the source buffer.
\item \npsafter. The null positions that remain unchanged in the destination buffer because they appear after $e$.
	This is defined by 
	    \text{$\rnps(\nps_x,\size{}_x, 0, \npfromy_{xy}, \geq  )$} in
	Figure~\ref{fig:flow.fun}.
\end{enumerate}

The difference between the \strcat and \strcpy functions is evident from the definition in 
	Figure~\ref{fig:flow.fun}---for \strcpy, the starting point of copying is 
$\start_{x}$
whereas for \strcat it is $\strend_{x}$. Further, their length limited versions choose the
minimum between the null distance and the length provided.

\subsubsection*{Set of Null Positions for Memory Copy}

We view \text{$\memcpy(x,y,m)$} as a special case of \text{$\strcpy(x,y)$}. The main difference is that 
\text{$\strcpy(x,y)$} copies from $y$ only upto the first null
character whereas \text{$\memcpy(x,y,m)$} copies $m$ characters from $y$. Thus the main change is 
in computing \npsfromy instead of \text{$\npat = \{\npfromy\}$}. Former includes multiple position whereas the latter 
computes a single position. Thus we compute
\begin{enumerate}
\item \npsfromy. This identifies all null character positions that lie between \text{$\start_y$} and 
      \text{$\start_y + m$} in the source buffer. These should be shifted by the start position 
        in the destination buffer. This is computed by the intersection of
	  \text{$
	  \rnps(\nps_y,\size{}_x,\sdist_{xy}, \start_y, \geq )
	 $}
	with 
	  \text{$
	\rnps(\nps_y,\size{}_x,\sdist_{xy}, \start_y + m, \leq ) 
	$} where \sdist represents the distance by which the null positions should be shifted. 
\item  \oflow. We must include $\infty$ to indicate overflow if
	\begin{itemize}
	\item \text{$m+\start_{x} > \size{}_x$} (the write operation crosses the destination boundary), or
	\item \text{$m+\start_{y} > \size{}_y$} (the read operation crosses the source boundary).
	\end{itemize}
	This is easily achieved by computing the following set differences
	\text{$\{m\oplus_{\size{}_x}\start_{x}  \} - \{m+\start_{x} \}$} and 
	\text{$\{m\oplus_{\size{}_y}\start_{y}  \} - \{m+\start_{y} \}$}.
\end{enumerate}

\subsection{Overflow Detection}
\label{sec:oflow.detection}

Computation of data flow values $\bin_n/\bout_n$ detects a buffer overflow at a program point at which a buffer is written into
by introducing $\infty$ in the \nps set. Note that this is a conservative conclusion and may well be a false positive.
Our analysis is sound when it concludes the absence of buffer overflow.

For a program statement that merely reads a buffer (eg. call to \strlen), the $\bin_n/\bout_n$ values remain unmodified. 
Detecting a potential buffer overflow for a read using a pointer is trivially achieved by 
\begin{itemize}[label=$-$]
\item Checking the buffer identity. If it is $b_\infty$, there is a potential buffer overflow.
\item Checking the buffer size. If it is $\infty$, there is a potential buffer overflow.
\item Checking the offset of the pointer. If it is $\infty$, there is a potential buffer overflow.
\item Otherwise, we compute the \strend of the pointer being read. If it is $\infty$, there is a potential buffer overflow.
\end{itemize}

\section{Running Example Revisited}
\label{sec:exmp}

This section shows the application of our analysis to 
the example in Figure~\ref{fig:exmp.1} and describes how buffer overflows can be detected.

\subsection{Computing the Set of Null Positions}

It is clear from Figure~\ref{fig:flow.fun} that the only non-obvious computation in this analysis is the computation
of $R_n$.  We illustrate it for our running example.  The relevant
maps and the values of the default extractor functions have already been provided 
in Figure~\ref{fig:exmp.1}. Observe that for the first two calls, buffer $b_1$ serves both as source and destination whereas
for the third call, buffer $b_0$ is the source and $b_1$ is the destination.

\begin{itemize}
\item After the first call to \text{$\strcat(z,y)$} in the example in Figure \ref{fig:exmp.1}, we have
\begin{align*}
	\npfromy &= \nullpos \oplus_{14} \copypos 
	\\ & = (\strend(\bm,\bp,y,b_1) - \start(\bp,y,b_1)) \oplus_{14} \strend(\bm,\bp, z, b_1)
	\\
	&= \left(\glb(\rnps(\{3,7,13\},14,0,4, \geq)) - \start(\bp,y,b_1)\right) 
	\\
	&\phantom{=}\;\; \oplus_{14} 
	\\
	&\phantom{=}\;\; (\glb(\rnps(\{3,7,13\},14,0,6, \geq)) 
	\\
	& = (\glb(\{ 7\oplus_{14} 0, 13 \oplus_{14} 0 \}) - 4) \oplus_{14}
	    (\glb(\{ 7\oplus_{14} 0, 13 \oplus_{14} 0 \}))
	\\
	& = (\glb(\{ 7, 13 \}) - 4) \oplus_{14} 
		(\glb(\{ 7, 13 \}) ) 
	\\
	& = (7 - 4) \oplus_{14} 7 = 10
\end{align*}
 The computation of relevant null positions (i.e. $R_n$) is as follows:
\begin{align*}
R_n(\bm,\bp,b_1) & = 
	\rnps(\{3,7,13\}, 14, 0, 6, <) 
        \cup  \{ 10 \}
        \cup 
	\rnps(\{3,7,13\}, 14, 0, 10, \geq) 
	\\
	&= 
	\{ 3 \oplus_{14} 0 \} 
	\cup \{ 10 \}
	\cup \{ 13 \oplus_{14} 0  \}  
	\\
&= 
	\{ 3 \} \cup  \{ 10 \} \cup \{ 13 \} = \{ 3, 10, 13 \} 
\end{align*}
\item After the second call to \strcat in the example, we have
\begin{align*}
	\npfromy &= \nullpos \oplus_{14} \copypos 
		\\
	& = (\strend(\bm,\bp,y,b_1) - \start(\bp,y,b_1)) \oplus_{14} \strend(\bm,\bp, z, b_1)
	\\
	&= \left(\glb(\rnps(\{3,10,13\},14,0,4, \geq)) - \start(\bp,y,b_1)\right) 
	\\
	&\phantom{=}\;\; \oplus_{14} 
	\\
	&\phantom{=}\;\; (\glb(\rnps(\{3,10,13\},14,0,6, \geq)) 
	\\
	& = (\glb(\{ 10\oplus_{14} 0, 13 \oplus_{14} 0 \}) - 4) \oplus_{14}
	    (\glb(\{ 10\oplus_{14} 0, 13 \oplus_{14} 0 \}))
	\\
	& = (\glb(\{ 10, 13 \}) - 4) \oplus_{14} 
		(\glb(\{ 10, 13 \}) ) 
	\\
	& = (10 - 4) \oplus_{14} 10 = \infty
\end{align*}
$R_n$ computation is as follows.
\begin{align*}
R_n(\bm,\bp,b_1) & = 
	\rnps(\{3,7,13\}, 14, 0, 6, <) \cup  \{ \infty \}
        \cup \rnps(\{3,7,13\}, 14, 0, \infty, \geq) 
	\\
	&= 
	\{ 3 \oplus_{14} 0 \} 
	\cup \{ \infty \}
	\cup \emptyset  
	\\
&= 
	\{ 3 \} \cup \{ \infty \} = \{ 3, \infty \} 
\end{align*}
\item The third call \text{$\strcpy(y,w)$} involves $b_0$ as the source buffer and $b_1$ as the destination buffer.
\begin{align*}
	\npfromy &= \nullpos \oplus_{14} \copypos 
	\\
	& = (\strend(\bm,\bp,w,b_0) - \start(\bp,w,b_0)) \oplus_{14} \start(\bm,\bp, y, b_1)
	\\
	&= \left(\glb(\rnps(\{10\},10,0,0, \geq)) - 0\right) \oplus_{14} 4
	\\
	& = \glb(\{ 10\oplus_{10} 0 \})  \oplus_{14} 4
	\\
	& = \glb(\{ 10 \}) \oplus_{14} 4
	\\
	& = 10  \oplus_{14} 4 = 14
\end{align*}
$R_n$ computation is as follows.
\begin{align*}
R_n(\bm,\bp,b_1) & = 
	\rnps(\{3,\infty\}, 14, 0, 4, <) \cup  \{ 14 \}
        \cup \rnps(\{3,\infty\}, 14, 0, 14, \geq) 
	\\
	&= 
	\{ 3 \oplus_{14} 0 \} 
	\cup \{ 14 \}
	\cup \{ \infty \} 
	\\
&= 
	\{ 3 \} \cup \{14 \}\cup \{ \infty \} = \{ 3, 14, \infty \} 
\end{align*}
Note that the $\infty$ has not been generated in the computation of \nullpos, it has been carried forward from the 
previous \nps set. In other words, this call does not cause an overflow, the presence of $\infty$ in $R_n$ indicates 
that an overflow has occurred in the buffer earlier.
\end{itemize}

\subsection{Overflow Detection}

It is clear from Section~\ref{sec:oflow.detection}
that checking overflow is trivial except in the case where \strend is to be used. We illustrate 
such uses for our running example in the following cases.
\begin{enumerate}
\item Assume that a statement \text{\tt v = x+14} is added to our running example.
      Then \text{$\start(\bp,v, b_1) = 14$}.  For simplicity, assume that 
      a call to $\strlen(v)$ occurs before the first call to \strcat. We compute
     \begin{align*}
     \strend(\bm,\bp,v,b_1) & = 
	\glb(\rnps(\{3,7,13\},14,0,14,\geq))
	\\
	&= \glb(\emptyset) = \infty
     \end{align*}
     Thus $\strlen(v)$ may cause a buffer overflow. In this case, this is not a false negative but a certain buffer overflow. However,
     our analysis does not have any means of distinguishing between a certain overflow and a false negative. 

\item For $\strlen(y)$ (before the first call to \strcat in the same example), we compute
     \begin{align*}
     \strend(\bm,\bp,y,b_1) & = 
	\glb(\rnps(\{3,7,13\},14,0,4,\geq))
	\\
	&= \glb(\{7 \oplus_{14} 0, 13\oplus_{14} 0\})
	\\
	&= \glb(\{7, 13\}) = 7
     \end{align*}
     The result indicates that the read will encounter a null character in the buffer at position 
     7 and hence there is no buffer overflow.
\item For $\strlen(y)$ (after the first call to \strcat in the same example), we compute
     \begin{align*}
     \strend(\bm,\bp,y,b_1) & = 
	\glb(\rnps(\{3,10,13\},14,0,4,\geq))
	\\
	&= \glb(\{10 \oplus_{14} 0, 13\oplus_{14} 0\})
	\\
	&= \glb(\{10, 13\}) = 10
     \end{align*}
      The result indicates that the read will encounter a null character in the buffer at position 10 
      and hence there is no buffer overflow.
\item For $\strlen(y)$ (after the second call to \strcat in the same example), we compute
     \begin{align*}
     \strend(\bm,\bp,y,b_1) & = 
	\glb(\rnps(\{3,\infty\},14,0,4,\geq))
	\\
	&= \glb(\emptyset) = \infty
     \end{align*}
The result indicates that the read may not find any null character in the buffer and hence there is a potential buffer overflow.
In this case also, this is a certain buffer overflow but we have no way of concluding so.
\end{enumerate}

\section{Extensions for Handling Multiple Pointee Buffers of a Pointer}
\label{sec:extensions}

In this section we show how our formulation can be extended to allow a pointer to point to multiple buffers at 
a program point. This requires a change 
in the lattice $B$, in the flow function \updatebuf, and in the computation of extractor function $R_n$ of Figure
\ref{fig:flow.fun}. These changes allow some relaxation in the assumptions about program model.

\subsection{Changes in the Lattices}

Given \text{$b\neq b'$} and 
\text{$i\neq j$}, we allow the coexistence of triples \text{$(x,b,i)$} and \text{$(x,b',j)$} in a \text{$\bp \in \bpt$}.
However, multiple triples of the kind \text{$(x,b,i)$} and \text{$(x,b,j)$} are still prohibited.

This is achieved by redefining \bpt for the lattice \text{$(2^\bpt, \bO)$}
as \text{$\bpt = \left(\pts \times\biL\right)  \mapsto \sizeL$}. Observe that now we use 
\text{$\biL = \{ b_i \mid i \in \nds \}$} and not \biinfL because now we do not
need the fictitious ``undefined'' buffer $b_\infty$. Also, because of the cross product 
\text{$\pts \times\biL$}, \text{$\pointee(\bp,x)$} now returns a set of buffers rather than a single buffer.

The meet operation \bM is simplified as follows:
\begin{align*}
\forall \, \bp,\bp' \in B\!: \;\;
\bp \,\bM \bp'
	& = 
	   \left\{
		\left( x, b, k \,\sizeM k' \right)
		\;\middle| \;
		\left( x, b, k \right) \in \bp, 
		\left( x, b, k' \right) \in \bp'  
	   \right\}
\end{align*}
Exclusion of $b_\infty$ also leads to a change in the boundary information in the data flow equation for \bin in which 
the triple \text{$\left(x, b_\infty, \infty \right)$} is replaced by 
the triple \text{$\left(x, b, \infty \right)$} for \Start{}.

\subsection{Changes in Extractor Function $R_n$}

When we have multiple source destination buffers in a string operation, we use the following approximation
by combining the effects of all these buffers: We use
the longest string among all destination buffers,
the smallest size among all source buffers, and
the farthest position of copying among all source buffers.
These three changes are reflected in our formulation by computing a single approximate $R_n$ 
by changing the terms appearing in Figure~\ref{fig:flow.fun}.

We describe the changes in $R_n$ computation for string operations.
\begin{itemize}
\item {\em Computing the longest string among all destination buffers.}
      We compute $\nullpos_y$ separately for each pointee buffer of $y$ and 
      take the largest value.
\item {\em Computing the smallest size among all source buffers.}
      We take the smallest value of $\size{}_x$ among all pointee buffers of $x$ for computing
      $\npfromy_{xy}$.

\item {\em Computing the farthest position of copying among all source buffers.}
      We take the largest value of $\copypos_{x}$ among all pointee buffers of $x$.

\item {\em Computing $R_n$.} This involves the following changes.
      \begin{itemize}
      \item Computing $\npsbefore_{xy}$ and $\npsafter_{xy}$.
        	We take the smallest value of $\size{}_x$ and the largest value of $\npfromy_{xy}$. 
      \item Computing $\npat_{xy}$.
      		We take the largest value of $\npfromy_{xy}$.
      \end{itemize}
\end{itemize}

Observe that these changes are declarative in the sense that they basically involve ranging over the buffers and
computing the maximum or the minimum. Hence a stateless formulation of these changes is easy to write.

Similar changes can be defined for the \memcpy operation.

\subsection{Changes in Flow Function \updatebuf}

With the possibility of a pointer pointing to multiple buffers, $D_n$ now becomes a set of buffers
and $K_n$ is computed separately for each buffer. However, $R_n$ is common because it is an approximation
of all source buffers. A buffer mapping \bm should accumulate updates in all source buffers in 
$D_n$. In other words, when a buffer $b \in D_n$ is updated, the resulting mapping $\bm'$ should be
passed on to \updatebuf for updating some other buffer $b' \in D_n$.
This is achieved by defining \updatebuf recursively and passing $D_n$ as the value of argument $X$ for the
top level call to \updatebuf. 
\begin{align*}
\!\!\!
\!\!\!
\!
\updatebuf(\bm,\bp,n,X)
	&= 
	\begin{cases}
	\updatebuf\left(
	\bm\left[b \mapsto ( 
			K_n(\bm,\bp,b),
			R_n(\bm,\bp)
                             )\right], \;
			\bpt, \;
			n, X \!- \!\{ b \}
	\right)
		& \renewcommand{\arraystretch}{.9}
			\begin{array}{@{}l}
			b \in X
		  \end{array}
		\\
	\bm 
		& \text{Otherwise}
		\rule{0em}{1.5em}
	\end{cases}
\end{align*}
In each recursive call, we accumulate the updates and the set $X$ becomes smaller. When 
all updates are accumulated in \bm, $X$ becomes $\emptyset$. 

\subsection{Relaxation in Program Model}

With the extensions for handling multiple buffers of a pointer at a program point in place,
now we do not need to ignore the back edges and we may be able to handle some cases where
range information is not available. However, this computation may be expensive as a loop
may be iterated many times depending upon the increment in the values of the pointers or
array indices.

\section{Related Work}
\label{sec:related}

Buffer overflow detection and mitigation has been an important concern for a long time. 
The problem is compounded by the idiosyncrasies of C string operations~\cite{Wagner00afirst} and
non-trivial semantics~\cite{Dor:2003:CTR:781131.781149}.  There is an 
abundance of literature on the topic and many tools have been created to address the concern; we cite only a few 
references as a starting point~\cite{%
citeulike:6345464,%
buf.study,%
Lhee:2003:BOF:781669.781672,%
Shahriar:2010:MBO:2441114.2441116,%
Zitser:2004:TSA:1029894.1029911%
}.

A popular approach of addressing this concern has been to perform
run time checks using 
code instrumentation, or 
expanded representation of pointers and buffers to store bookkeeping information (effected through compilers and changed library 
function)~\cite{%
Dhurjati:2006:BAB:1134285.1134309,%
Jones97backwards-compatiblebounds,%
Ruwase04apractical%
}.
The known static analysis methods are characterized by some combination of the following features:
\begin{itemize}
\item User annotations in the program~\cite{Dor:2003:CTR:781131.781149,Larochelle:2001:SDL:1251327.1251341} in the form of contracts
      or assertions.
\item Range analysis~\cite{Li:2010:PES:1882291.1882338,Wagner00afirst} storing the lower and the upper limits of the
      buffers. 
\item Integer linear programming~\cite{Dor:2003:CTR:781131.781149,Ganapathy:2003:BOD:948109.948155,Wagner00afirst} to solve constraints
to compute the ranges.
\item Symbolic computation~\cite{Li:2010:PES:1882291.1882338} to store the ranges in terms of expressions rather than in terms of integers.
\end{itemize}
Most approaches set up constraints flow insensitively (or use flow insensitive pointer analysis) 
except~\cite{Li:2010:PES:1882291.1882338} which does path sensitive analysis by storing relevant path 
predicates.\footnote{It avoids a
combinatorial explosion by a sparse demand driven computation that uses transitive data and control dependences.}
What is common to all these approaches is that their formulations are stateful algorithms; 
none of them have a stateless formalization amenable to reasoning and automatic construction of analysers.

In terms of modelling, most approaches do not separate pointers and buffers as we do; they need to explicitly discover
aliasing between buffers to record the effects of one change through a pointer, into the buffer of another pointer. 
In this sense our modelling is closer to the modelling 
for ``intended referent''~\cite{Jones97backwards-compatiblebounds}. However they store the information for
run time checking, we use it for static analysis. Besides, unlike them, we do not treat a string as a buffer and instead 
store a set of null positions within a buffer. Most approaches treat a string as a buffer and hence store a single length
with a buffer.

\section{Conclusions}
\label{sec:conclusions}

Modelling buffers and buffer overflows in terms of 
\begin{itemize}
\item mappings between pointers and buffers whose addresses they hold,
\item buffer information storing size and sets of positions of null characters, and 
\item defining functions to compute these models
\end{itemize}
makes it possible to formulate buffer overflow analysis as a data
flow analysis of C programs. A key facilitator of this is an emphasis
on stateless formalizations of analyses in terms of lattice valued
functions and relations. A stateful formulation combines features
through side effects recorded in states thereby raising a natural
requirement of C/C++/Java code to be written to complement a partially
high level specification so that a generator can generate the analyser.

Banishing states from formulations enable higher levels of abstraction.
The resulting conciseness, together with higher levels of abstraction,
makes the formulations amenable to human reasoning. Further, such
formulations allow a generator to check the specifications, combine
their features freely, and decide how and where to introduce states in
the generated code.

This paper does not claim to define the best buffer overflow
analysis; it may well be possible to devise better static
approximations. In particular, 
storing ``out of bound'' null positions~\cite{Ruwase04apractical} may eliminate many false positives.
It may also be interesting to explore the possibility of using symbolic
      expressions~\cite{Li:2010:PES:1882291.1882338} for modelling the set
     of null character positions. 

So long as we have a well defined static
approximation, we believe that a stateless mathematical formulation
leading to high level specifications amenable to automatic construction
of analysers is possible.

\section*{Acknowledgements}
Dipali Bhutada, Kalyani Zope, Swati Jaiswal, and Vini Kanvar
 provided useful feedback on the drafts of this paper.

\bibliography{bof}

\begin{thebibliography}{10}

\bibitem{citeulike:6345464}
C.~Cowan, F.~Wagle, Calton Pu, S.~Beattie, and J.~Walpole.
\newblock {Buffer overflows: attacks and defenses for the vulnerability of the
  decade}.
\newblock In {\em DARPA Information Survivability Conference and Exposition,
  2000. DISCEX '00. Proceedings}, volume~2, 2000.

\bibitem{Dhurjati:2006:BAB:1134285.1134309}
Dinakar Dhurjati and Vikram Adve.
\newblock Backwards-compatible array bounds checking for c with very low
  overhead.
\newblock In {\em Proceedings of the 28th International Conference on Software
  Engineering}, ICSE '06, pages 162--171, New York, NY, USA, 2006. ACM.

\bibitem{Dor:2003:CTR:781131.781149}
Nurit Dor, Michael Rodeh, and Mooly Sagiv.
\newblock Cssv: Towards a realistic tool for statically detecting all buffer
  overflows in c.
\newblock In {\em Proceedings of the ACM SIGPLAN 2003 Conference on Programming
  Language Design and Implementation}, PLDI '03, pages 155--167, New York, NY,
  USA, 2003. ACM.

\bibitem{buf.study}
Pierre-Alian Fayolle and Vincent Glaume.
\newblock A buffer overflow study: Attacks and defenses.
\newblock Unpublished manuscipt available at {\tt
  http://repo.hackerzvoice.net/depot\_ouah/bofstd.pdf}. (Last accessed on 16
  Dec 2014), 2002.

\bibitem{Ganapathy:2003:BOD:948109.948155}
Vinod Ganapathy, Somesh Jha, David Chandler, David Melski, and David Vitek.
\newblock Buffer overrun detection using linear programming and static
  analysis.
\newblock In {\em Proceedings of the 10th ACM Conference on Computer and
  Communications Security}, CCS '03, pages 345--354, New York, NY, USA, 2003.
  ACM.

\bibitem{Jones97backwards-compatiblebounds}
Richard W~M Jones, Paul H~J Kelly, Most C, and Uncaught Errors.
\newblock Backwards-compatible bounds checking for arrays and pointers in c
  programs.
\newblock In {\em in Distributed Enterprise Applications. HP Labs Tech Report},
  pages 255--283, 1997.

\bibitem{Khedker.UP.Karkare.B:2008:Efficiency-Precision-Simplicity}
U.~P. Khedker and B.~Karkare.
\newblock Efficiency, precision, simplicity, and generality in interprocedural
  data flow analysis: Resurrecting the classical call strings method.
\newblock In {\em Proceedings of the International Conference on Compiler
  Construction}, pages 213--228. Springer-Verlag, 2008.

\bibitem{Khedker.UP.Sanyal.A.Karkare.B:2009:Data-Flow-Analysis}
U.~P. Khedker, A.~Sanyal, and B.~Karkare.
\newblock {\em Data Flow Analysis: Theory and Practice}.
\newblock CRC Press (Taylor and Francis Group), 2009.
\newblock (Under publication).

\bibitem{Larochelle:2001:SDL:1251327.1251341}
David Larochelle and David Evans.
\newblock Statically detecting likely buffer overflow vulnerabilities.
\newblock In {\em Proceedings of the 10th Conference on USENIX Security
  Symposium - Volume 10}, SSYM'01, Berkeley, CA, USA, 2001. USENIX Association.

\bibitem{Lhee:2003:BOF:781669.781672}
Kyung-Suk Lhee and Steve~J. Chapin.
\newblock Buffer overflow and format string overflow vulnerabilities.
\newblock {\em Softw. Pract. Exper.}, 33(5):423--460, April 2003.

\bibitem{Li:2010:PES:1882291.1882338}
Lian Li, Cristina Cifuentes, and Nathan Keynes.
\newblock Practical and effective symbolic analysis for buffer overflow
  detection.
\newblock In {\em Proceedings of the Eighteenth ACM SIGSOFT International
  Symposium on Foundations of Software Engineering}, FSE '10, pages 317--326,
  New York, NY, USA, 2010. ACM.

\bibitem{Padhye:2013:IDF:2487568.2487569}
Rohan Padhye and Uday~P. Khedker.
\newblock Interprocedural data flow analysis in soot using value contexts.
\newblock In {\em Proceedings of the 2Nd ACM SIGPLAN International Workshop on
  State Of the Art in Java Program Analysis}, SOAP '13, pages 31--36, New York,
  NY, USA, 2013. ACM.

\bibitem{Ruwase04apractical}
Olatunji Ruwase and Monica~S. Lam.
\newblock A practical dynamic buffer overflow detector.
\newblock In {\em In Proceedings of the 11th Annual Network and Distributed
  System Security Symposium}, pages 159--169, 2004.

\bibitem{Shahriar:2010:MBO:2441114.2441116}
Hossain Shahriar and Mohammad Zulkernine.
\newblock Monitoring buffer overflow attacks: A perennial task.
\newblock {\em Int. J. Secur. Softw. Eng.}, 1(3):18--40, July 2010.

\bibitem{Wagner00afirst}
David Wagner, Jeffrey~S. Foster, Eric~A. Brewer, and Alexander Aiken.
\newblock A first step towards automated detection of buffer overrun
  vulnerabilities.
\newblock In {\em In Network and Distributed System Security Symposium}, pages
  3--17, 2000.

\bibitem{Zitser:2004:TSA:1029894.1029911}
Misha Zitser, Richard Lippmann, and Tim Leek.
\newblock Testing static analysis tools using exploitable buffer overflows from
  open source code.
\newblock In {\em Proceedings of the 12th ACM SIGSOFT Twelfth International
  Symposium on Foundations of Software Engineering}, SIGSOFT '04/FSE-12, pages
  97--106, New York, NY, USA, 2004. ACM.

\end{thebibliography}

\appendix

\section{Modelling Other Statements}
\label{app:stmt.modelling}

We show how most other statements can be modelled in terms of the core statements of
our formulation. 
\begin{itemize}
\item Assignment of string literals (eg. a statement \text{$x = ``Hello \; world";$})
      is modelled as a sequence of two statements \text{$x = \malloc(k);
      x[k+1] =\!'\backslash{}0';$} where $k$ represents the length of the string literal.
\item Reallocation of memory to resize a buffer using \text{$x = \realloc(y,k)$} is
      modelled as a sequence of two statements \text{$x = \malloc(k);
      \strcpy(x,y)$}. 
\item Statements \text{$*(x+i) = \,'\backslash{}0'$} and
       \text{$*(x+i) =\, '\!c'$} are equivalent to \text{$x[i] =
       \,'\backslash{}0'$} and \text{$x[i] = \,'\!c'$} respectively.
\item A declaration of an uninitialized array (eg. \text{$char \; x[k];$})
      is modelled as \text{$\malloc(k)$} statement.
\item A declaration of an initialized array (eg. \text{$char \; x[k] =  \{ \ldots \};$})
      is modelled as a sequence of two statements \text{$x = \malloc(k);
      x[k+1] =\!'\backslash{}0';$} where $k$ represents the number of elements in the array.
\item The following string functions are modelled in terms of \strlen 
      for the purpose of buffer overflow analysis:
      \text{\sf\em strcmp$(x,y)$},
      \text{\sf\em strncmp$(x,y)$},
      \text{\sf\em strchr$(x,y)$}, and
      \text{\sf\em strrchr$(x,y)$}. The function
      \text{\sf\em strstr$(x,y)$} finds the first occurrence of string $y$ in string $x$ and
      is viewed as a combination of 
      \text{$\strlen(x)$} and \text{$\strlen(y)$} for this analysis. A call to
      \text{\sf\em strtok$(x,y)$} is similarly modelled in terms of 
      \text{$\strlen(x)$} and \text{$\strlen(y)$}. 
   
      Some of these functions either return a buffer pointer containing a possible
      substring (or a collection of substrings). Their lengths are dynamic and hence
      are not amenable to static analysis. Hence, we view the
	functions returning pointers to buffers with dynamic lengths, as
              functions creating buffers of size $\infty$. 
\item The remaining memory handling functions can also be similarly modelled and statically approximated in case their
      result depends on a dynamic value.
\end{itemize}

We do not rule out the possibility of better static approximations and expect them to be handled in a similar manner provided they
can be defined precisely in terms of stateless functions.

\end{document}